# Anomalous impact of thermal fluctuations on spin transfer torque induced ferrimagnetic switching

Zhengping Yuan[1,*], Jingwei Long[1,*], Zhengde Xu[1], Yue Xin[1], Lihua An[1], Jie Ren[1], Xue Zhang[1], Yumeng Yang[1,2] and Zhifeng Zhu[1,2,†]

[1]School of Information Science and Technology, ShanghaiTech University, Shanghai, China 201210

[2]Shanghai Engineering Research Center of Energy Efficient and Custom AI IC, Shanghai, China 201210

**Abstract**

The dynamics of a spin torque driven ferrimagnetic (FiM) system is investigated using the two-sublattice macrospin model. We demonstrate an ultrafast switching in the picosecond range. However, we find that the excessive current leads to the magnetic oscillation. Therefore, faster switching cannot be achieved by unlimitedly increasing the current. By systematically studying the impact of thermal fluctuations, we find the dynamics of FiMs can also be distinguished into the precessional region, the thermally activated region, and the cross-over region. However, in the precessional region, there is a significant deviation between FiM and ferromagnet (FM), i.e., the FM is insensitive to thermal fluctuations since its switching is only determined by the amount of net charge. In contrast, we find that the thermal effect is pronounced even a very short current pulse is applied to the FiM. We attribute this anomalous effect to the complex relation between the anisotropy and overdrive current. By controlling the magnetic anisotropy, we demonstrate that the FiM can also be configured to be insensitive to thermal fluctuations. This controllable thermal property makes the FiM promising in many emerging applications such as the implementation of tunable activation functions in the neuromorphic computing.

**Introduction**

Moore's Law has accurately predicted the development of integrated circuit technology since its inception in 1965.[1] However, due to the limitations such as the cost, heat generation, and Heisenberg uncertainty, integrating more transistors per unit area is becoming increasingly difficult with the existing integrated circuit processes.[2-4] Various alternative routes are considered feasible,[5-7] among which spintronics is viewed as one of the solutions in the post-Moore era and has attracted enormous attention. Spintronic devices are favored due to their non-volatility, high energy efficiency, and compatibility with CMOS processes.[5] Early studies have shown that the magnetization can be manipulated by applying an external magnetic field.[8, 9] Furthermore, the discovery of giant magnetoresistance (GMR) effect makes it feasible to read the state of the magnetic moment through the resistance. This further demonstrates the potential of spintronic devices for practical applications.[10] However, the magnetic field driven devices suffer from difficulties in the scaling down due to the large driving current.[11] Benefiting from the discovery of spin transfer torque (STT),[12, 13] it became possible to use spin polarized currents to manipulate the magnetic moment, which enables an all-electric control of spintronic devices. STT-driven magnetic tunnel junctions (MTJs) using ferromagnetic (FM) materials have been extensively studied. Their dynamics of FM has been comprehensively explained.[14-16] The commercial products of STT magnetic random-access memory (MRAM) are already appeared on the market.[17-19] Compared to the FM, the antiferromagnet (AFM) possesses unique dynamic characteristics.[20-22] AFM has been demonstrated to exhibit ultrafast dynamics.[23] However, the manipulation of AFM order is complicated as a result of its zero net magnetic moment.[24, 25] Similar to AFMs, ferrimagnets (FiMs) composed of antiferromagnetically coupled magnetic moments also possess large exchange interactions. Therefore, FiMs are also promising in achieving ultrafast magnetic dynamics. More importantly, their non-zero net magnetic moments allow easier electrical manipulation.[26-29] Commonly used FiM materials include alloys of transition metals (TM) and rare earth (RE) elements, YIG, and so on.[27]

Compared to FMs, FiM-based spintronics exhibit interesting properties already witnessed in ultrafast spin dynamics, as well as efficient electrical operations. Different methods have been implemented to explain the dynamics and compensation point-related properties, for example, one can calculate the effective parameters and then substitute them into the existing formula derived for FMs.[30] Additionally, the atomistic spin simulation is used in the study of temperature dependence of magnetization in FiM.[31] Alternatively, the Landau-Lifshitz-Bloch (LLB) equation can be used to take into account the longitudinal variation of magnetization.[32] Furthermore, deterministic FiM magnetic moment switching is feasible by picosecond electrical heating to excite conduction electrons in magnetic metals with charge-current pulses, without using spin torque.[33] In addition, Joule heating is considered to assist magnetic moment switching.[34] Moreover, the current-induced magnetization reversal in GdFeCo has strong random fluctuations in the time required for magnetic domain wall nucleation, and the switching process is thermally activated.[35] Although these studies can explain the switching of FiM, the switching process of FiM under the thermal fluctuations has not been fully understood. Additionally, magnetic anisotropy in FM systems is considered to be related to the critical current density and thermal stability.[36] Meanwhile, in FiM, anisotropy has been observed to have drastic influence on thermal stability[37] and thermally related effects (such as the longitudinal spin Seebeck effect[38]). Therefore, it is reasonable to question the effect of anisotropy on the sensitivity of FiM to thermal fluctuations. Noting that FiM is sensitive to thermal effects and the influence of thermal fluctuations on the device cannot be neglected in practical applications, a comprehensive understanding of the thermal fluctuations induced effects in the FiM systems is indispensable.

In this paper, we investigate the dynamics of STT-driven FiM considering thermal fluctuations with the two-sublattice macrospin model. Firstly, the switching time of the FiM/MgO/FM MTJ is studied, revealing its ultrafast dynamics in the picosecond region. Furthermore, the switching probability ($P_{SW}$) under thermal fluctuations is investigated, and we find that, in the precessional region, FiM is sensitive to the thermal effect. This is in stark contrast to $P_{SW}$ of FM that is only determined by the amount of net charge. In order to reveal the underlying physical mechanisms, the relationship between $P_{SW}$ and the net charge is investigated. We find that this anomalous impact of thermal fluctuations on FiM is attributed to the complicated relation between anisotropy and the overdrive current, which is consistent with the hypothesis that magnetic anisotropy would have a pronounced effect on the sensitivity of FiM to thermal fluctuations. This work provides guidance for the further understanding of FiM dynamics and supports the implementation of FiM-based spintronic devices.

**Methods**

The device structure of FM/MgO/FiM MTJ with perpendicular magnetic moments is shown in Fig. 1(a), in which GdFeCo is used as the free layer (FL), with a dimension of 50 nm×50 nm×0.6 nm, FM is used as the pinned layer (PL), and the spacer layer MgO is in the middle. In addition, the charge current ($J_C$) is applied along the **z** direction to generate STT. As shown in Fig. 1(b), the two-sublattice macrospin model is used. Different from the macrospin model with a single magnetic moment, here Gd and FeCo are regarded as two uniform sublattices, respectively. As a result, two coupled Landau-Lifshitz-Gilbert-Slonczewski (LLGS) equations[39] are used to describe the dynamics of the two magnetic moments,

$$\frac{\partial \mathbf{m}_1}{\partial t} = -\gamma_1 \mu_0 \mathbf{m}_1 \times \mathbf{H}_1 + \alpha_1 \mathbf{m}_1 \times \frac{\partial \mathbf{m}_1}{\partial t} - \gamma_1 \mu_0 \mathbf{m}_1 \times \left(\mathbf{m}_1 \times \mathbf{H}_1^{DLT}\right) \quad (1)$$

$$\frac{\partial \mathbf{m}_2}{\partial t} = -\gamma_2 \mu_0 \mathbf{m}_2 \times \mathbf{H}_2 + \alpha_2 \mathbf{m}_2 \times \frac{\partial \mathbf{m}_2}{\partial t} - \gamma_2 \mu_0 \mathbf{m}_2 \times \left(\mathbf{m}_2 \times \mathbf{H}_2^{DLT}\right) \quad (2)$$

where $\mathbf{H}_1 = \mathbf{H}_K^{Fe} + \mathbf{H}_{ex}^{Fe} + \mathbf{H}_{ext}$ and $\mathbf{H}_2 = \mathbf{H}_K^{Gd} + \mathbf{H}_{ex}^{Gd} + \mathbf{H}_{ext}$ are the effective fields, and $\gamma_1 = g_1 \mu_B/\hbar$ and $\gamma_2 = g_2 \mu_B/\hbar$ are gyromagnetic ratios. $\mu_0 \mathbf{H}_K^{Fe} = [0, 0, (2K_{u,Fe}/M_{s,Fe})m_{z,Fe}]$ and $\mu_0 \mathbf{H}_K^{Gd} = [0, 0, (2K_{u,Gd}/M_{s,Gd})m_{z,Gd}]$ are the anisotropy fields.

The exchange fields are $\mu_0\mathbf{H}_{ex}^{Fe} = -A_{ex}/M_{s,Fe}\mathbf{m}_{Gd}$ and $\mu_0\mathbf{H}_{ex}^{Gd} = -A_{ex}/M_{s,Gd}\mathbf{m}_{Fe}$, where $A_{ex}$ is the exchange constant. The third term on the right-hand side represents the damping-like torque (DLT) of STT with $\mu_0\mathbf{H}_1^{DLT} = (\eta\hbar J_C\boldsymbol{\sigma})/(2t_{FL}M_{s,Fe})$ and $\mu_0\mathbf{H}_2^{DLT} = (\eta\hbar J_C\boldsymbol{\sigma})/(2t_{FL}M_{s,Gd})$. The polarization of the spin current $\boldsymbol{\sigma}$ is [0, 0, 1], and the STT efficiency $\eta = P/(1 + P^2\cos(\theta))$ [40] with the polarization of the magnetic layer $P$. When thermal fluctuations are considered, they are included as random fields[41] $\mu_0\mathbf{H}_{therm1} = N(0,u_1)\hat{\mathbf{x}} + N(0,u_1)\hat{\mathbf{y}} + N(0,u_1)\hat{\mathbf{z}}$ and $\mu_0\mathbf{H}_{therm2} = N(0,u_2)\hat{\mathbf{x}} + N(0,u_2)\hat{\mathbf{y}} + N(0,u_2)\hat{\mathbf{z}}$, where $N(0,u_i)$ is the normal distribution. The standard deviations are $u_1 = \sqrt{2k_BT\alpha_1/(V_{FL}M_{s,Fe}\gamma_1(1+\alpha_1^2)\Delta t)}$ and $u_2 = \sqrt{2k_BT\alpha_2/(V_{FL}M_{s,Gd}\gamma_2(1+\alpha_2^2)\Delta t)}$ where $V_{FL}$ is the volume of FL and $\Delta t$ is the duration of thermal fluctuations. The parameters used in the simulation are as follows: $A_{ex} = 9\times10^6$ J/m³, $K_{u,Fe} = K_{u,Gd} = 1.2\times10^5$ J/m³, $\alpha_1 = 0.01$, $\alpha_2 = 0.02$, $g_1 = 2.2$, $g_2 = 2$, $P = 0.4$. In this work, we ignore Joule heating, meanwhile, a fixed temperature $T = 300$ K is set with $M_{s,Fe} = 1149$ kA/m and $M_{s,Gd} = 1012$ kA/m. Additionally, due to the existence of thermal fluctuations induced random fields, the switching of magnetic moments is random under the same pulse width ($\tau$) and $J_C$. Therefore, to calculate the switching probability ($P_{SW}$), 100 independent runs under the same $\tau$ and $J_C$ are performed, and $P_{SW}$ is calculated by dividing the number of switching events by the total number of runs.

**Results and Discussions**

  The time scale of magnetization switching has always attracted much attention,

especially for materials with antiferromagnetic coupling such as AFM and FiM, which have potential for ultrafast spintronic devices. When the thermal fluctuations are not considered, the temporal evolution of the magnetization of two sublattices at $J_C$ = $3\times10^{11}$ A/m$^2$ is shown in Fig. 2(a). During the switching process, $\mathbf{m}_1$ acts as the dominant sublattice since its saturation magnetization is larger. Driven by STT, $\mathbf{m}_1$ overcomes the magnetic anisotropy and finally aligns to the other direction of the easy axis, i.e., up to down switching. In the meanwhile, due to the existence of exchange interaction, $\mathbf{m}_2$ is switched from down to up. In addition, the tilting angle between the two sublattices during the switching process is shown in Fig. 2(b), and the two sublattices are not always antiparallel with the largest angle deviation of 2.39° at $t$ = 29.6 ps. Due to the existence of exchange coupling torque induced by antiferromagnetic exchange coupling, the angle deviation contributes to the ultrafast switching observed here.[42] This angle deviation is different from the common assumptions that the two sublattices of FiM are always antiparallel during the switching process.[28, 43]

By applying different $J_C$, the dynamics of the sublattices under the action of STT are obtained, from which one can determine the boundary of switching and non-switching, i.e., the critical current density $J_{C0}$ = 2.26×10$^{11}$ A/m$^2$. Fig. 2(c) shows the relationship between $J_C$ and the switching time ($t_{SW}$), which is defined as the time for the $\mathbf{m}_1$ to move from the initial position to the critical position ($m_{1,z}$= 0). The larger $J_C$, the larger the STT, which will make the switching faster. However, when $J_C$ is greater than 2.5×10$^{11}$ A/m$^2$, increasing $J_C$ has little effect on reducing $t_{SW}$. It is worth noting that the applied current density of FiM-based MTJs is comparable to that of MTJs based on FM, nevertheless, the switching time of FiM is much shorter than that of FM. FM-based MTJ has a switching time in the nanosecond range,[44] nevertheless, FiM possesses an ultrafast switching in the picosecond range.

Furthermore, as shown in Fig. 2(d), if $J_C$ continues to be increased, FiM will enter the oscillation region.[45] Here, the current density corresponds to the threshold of oscillation is defined as $J_{osc}$. Therefore, it should be noted that shorter $t_{sw}$ cannot be obtained by continuously increasing $J_C$. This means that an appropriate current density between $J_{C0}$ and $J_{osc}$ is required to achieve deterministic switching of FiM and this range is relatively small compared to the whole current density range of FiM phase.

In practical applications, thermal fluctuations are not negligible and have a significant impact on the magnetization dynamics. The dynamics of FM under thermal fluctuations has been comprehensively investigated. Whether driven by STT or spin-orbit torque,[26-29, 46] the magnetic dynamics in FM devices can be distinguished into three regions: the precessional region, the thermally activated region, and the cross-over region in between. In the precessional region, the current densities corresponding to 50% switching probability ($J_{C,50}$) is inversely proportional to the pulse width ($\tau$). In this case, the switching of magnetic moment is driven by spin angular momentum, and the dynamics of magnetization is only determined by the STT. In contrast, $J_{C,50}$ in the thermally activated region is proportional to the logarithm of the pulse width, and it is even smaller than $J_{C0}$. This is due to the fact that the dynamics of magnetization is mainly dominated by thermal fluctuations, and the energy required for switching is mainly provided by thermal activation. In the cross-over region, the switching of the magnetic moment is due to the combined effect of thermal activation and angular momentum transfer, and the effects of thermal fluctuations and STT are comparable. Following the definition of boundary conditions in the FM, the corresponding ones in the FiM are summarized in Table I, where the ratio of $J_{C,50}$ and $J_{C0}$ is defined as the overdrive $i$. It is worth noting that the value of $i$ has a decisive influence on the magnetization dynamics and the factors that affect the range of $i$, such as the magnetic anisotropy, the exchange constant, will make the entire system exhibit different characteristics under the action of thermal fluctuations.

**Table I.** Summary of boundary conditions

| Region | Mechanism | Boundary | $J_{C,50}$ vs $\tau$ |
|---|---|---|---|
| The precessional region | Angular momentum transfer | $i \gg 1$ | $J_{C,50} \propto 1/\tau$ |
| The cross-over region | Angular momentum transfer and thermal activation | $i \approx 1$ | / |
| The thermally activated region | Thermal activation | $i \ll 1$ | $J_{C,50} \propto \ln(\tau)$ |

To further investigate the dynamics of FiM, thermal fluctuations are considered, and pulsed currents are used. The total pulse width includes the rise time $t_{\text{rise}} = 100$ fs, fall time $t_{\text{fall}} = 100$ fs and plateau time.[47] Fig. 3(a) shows the $P_{\text{SW}}$ of the FiM system versus $J_C$, where one can see that a larger $J_C$ is required to achieve the same $P_{\text{SW}}$ when the pulse width is small. Fig. 3(b) shows the relationship between $i$ and the inverse of the pulse width. With a shorter pulse width, the thermal fluctuations will provide less energy, consequently, a larger overdrive current is required to generate sufficient STT, which drives the magnetic moment motion. However, unlike the results of FM systems in the precessional region,[48] $i$ is not proportional to the inverse of the pulse width and a significant deviation is observed. As we have mentioned that the range of $i$ is affected by several parameters, including the magnetic anisotropy energy and exchange interaction. Noting that the magnetic anisotropy plays an extremely important role in the FiM dynamics. It is intuitive to investigate the impact of anisotropy on the sensitivity of FiM to thermal fluctuations. To further understand this deviation, systems with different magnetic anisotropy energy or different exchange constants are studied. Although the change in the exchange interaction affects $J_{C0}$, it has little effect on the range of $i$. The results are similar to Fig. 3(b), which still deviates from the linear trend in the FM. In contrast, the relationship between $i$ and $1/\tau$ in the system with a smaller magnetic anisotropy energy ($K_{u,Fe} = K_{u,Gd} = 3 \times 10^4$ J/m$^3$) agrees well with that of FMs. Here, we define the system with the smaller magnetic anisotropy energy as the system 2 (S2), and the system studied in Fig. 3(b) is defined as the system 1 (S1). The critical current density of S2 is $J_{C0} = 5.6 \times 10^9$ A/m$^2$, which is much smaller than that of S1. Fig. 3(c) shows the $P_{SW}$ versus $J_C$ for different pulse widths in S2, and the corresponding $i$ and $1/\tau$ is plotted in Fig. 3(d). Similar to the dynamics of FM, a linear relationship between $i$ and $1/\tau$ is observed, which indicates the switching of the S2 is dominated by STT. Meanwhile, comparing Fig. 3(b) and Fig. 3(d), under the same pulse width, $i$ of S2 is significantly larger than that of S1. In addition, in the thermally activated region, the overdrive $i$ is proportional to the logarithm of the pulse width, we plot the $i$-lg($\tau$) curves of S1 and S2 (see Appendix A), the $i$-lg($\tau$) curve of S1 is close to a straight line even under small pulse width. However, the $i$ of S2 under small $\tau$ is proportional to $1/\tau$. Additionally, comparing Fig. 3(b), Fig.3(d) and Fig. 5, the results demonstrate that when $i\approx 1$, both S1 and S2 deviate from the linear $i$-$1/\tau$ relation and $i$-$\lg(\tau)$ relation, proving that both systems are in the cross-over region between the precessional region and the thermally activated region. Therefore, we attribute the deviation from the linearity of the $i$-$1/\tau$ curve of S1 to the influence of thermal fluctuations. The mechanism responsible for these differences in the two systems needs to be further investigated.

According to the LLG ballistic transport theory,[48] the angular momentum required for the switching of magnetic moments is constant. In other words, in absence of thermal fluctuations, the net charge required to achieve the same $P_{SW}$ is fixed, due to the conservation of angular momentum. Therefore, $P_{SW}$-net charge relationship ought to be not dependent on the pulse width or the charge current density, separately. To further investigate the impact of thermal fluctuations on the FiM system, the net charges, which are represented by $(J_C - J_{C0})\tau$, corresponding to different $P_{SW}$ at different pulse widths are measured. As shown in Fig. 4(a), for S1, the distributions of net charge under different pulse widths do not overlap at all. In the meanwhile, the larger the pulse width, the less the net charge is required to achieve the same $P_{SW}$. It is apparent that the longer pulse makes the thermal activation provide more energy to the system. In this case, the results indicate that thermal activation provides a non-negligible contribution to switching compared to the angular momentum transfer provided by STT. Therefore, the system is significantly affected by thermal fluctuations even under small pulse width. In contrast, when the pulse widths are small, the distributions of net charge in S2 are nearly identical [see Fig. 4(b)], which indicates that the contribution of thermal activation to switching is negligible compared to the angular momentum transfer provided by STT. This demonstrates that the magnetization switching is only determined by the amount of net charge. Additionally, we calculated the stand deviation of the net charge required to achieve the same $P_{SW}$ for S1 and S2 (see Appendix B), and the stand deviations of S2 is much smaller than that of S1. The results show S2 is less affected by thermal fluctuations compared to the other system. It is worth noting that only the difference in anisotropy causes S1 and S2 to exhibit different thermal properties, and it is possible to change the anisotropy of FM to obtain similar unique thermal properties. As a comparison, we have simulated FM systems with different anisotropy and find that changing the anisotropy has little effect on the sensitivity of FM to thermal fluctuations (see Appendix C for details). In the precessional region, the $P_{SW}$ of FM is determined by the net charge. These demonstrate that the controllable sensitivity to thermal fluctuations we observed is unique to FiM.

Comparing the magnetic dynamics in S1 and S2, we find that the control of magnetic anisotropy allows the FiM system to behave differently in response to the thermal effect. This controllable thermal immunity enables the integration of multiple functions in a single device. As shown in Fig. 4(a), the curves of $P_{SW}$ are similar to the sigmoid activation functions used in the neuromorphic computing.[49] In S1, the slope and shift of curves are strongly affected by the pulse width, which can be used in the neural network where tunable activation functions are required.[50] In contrast, the curves in S2 [see Fig. 4(b)] are nearly identical for the pulse widths range from 10 to 100 ps. Therefore, S2 can be used in the neural network which requires a fixed activation function. The reconfiguration between S1 and S2 can be achieved electrically

using voltage-controlled magnetic anisotropy.[51-54]

Fig. 4(c) and Fig. 4(d) show the relationship between the net charge $(J_{C,50}-J_{C0})\tau$ and pulse width at 50% switching probability. For S1, $(J_{C,50}-J_{C0})\tau$ decreases sharply at all time scales. This indicates that the switching process is strongly affected by thermal activations. In contrast, the STT dominates the switching process in S2 at small $\tau$, resulting in an almost unchanged $(J_{C,50}-J_{C0})\tau$. These results demonstrate that S2 is close to ballistic transport, and it is less affected by thermal fluctuations. In order to quantitatively compare the effects of thermal fluctuations on S1 and S2, we introduce the concept of "net STT" (see Appendix D for details) and we have compared the net STT of Fe in S1 and S2. It is found that S2 has a larger net STT than S1 at $\tau$ from 100ps to 1ns, proving that S1 is more susceptible to thermal fluctuations than S2. The reason responsible for the different behaviors of S1 and S2 in the precessional region is that S1 cannot provide a sufficiently large value of overdrive $i$. Since the boundary of the precessional region is $i \gg 1$, as shown in Table I, a larger $i$ can ensure that the system is stable in the precessional region and is less affected by thermal fluctuations. Small overdrive of S1 makes it more susceptible to thermal fluctuations. In addition, under the same $\tau$, $i$ corresponding to S2 is obviously larger than that of S1, as shown in Fig. 3(b) and Fig. 3(d), which makes the switching in S2 is completely determined by STT in the precessional region.

**Conclusion**

In this paper, we used the two-sublattice macrospin model to investigate the dynamics of FiM. The results demonstrate that the switching time of FiM is in the picosecond range. However, the excessive current density will result in the oscillation of FiM. Therefore, faster switching cannot be obtained by unlimitedly increasing $J_C$. Furthermore, the impact of thermal fluctuations on switching is investigated, which significantly affects $P_{SW}$ at all time scales. Similar to the FM, FiM can also be distinguished into three regions, namely the precessional region, the thermally activated region, and the cross-over region. The boundary conditions are defined by the overdrive $i$. The range of $i$ has decisive influence on the dynamics of FiM, and it is greatly affected by the magnetic anisotropy. By controlling the anisotropy, FiM can exhibit controllable sensitivity to thermal fluctuations in the precessional region. This unique thermal property makes FiM promising for the applications such as the neuromorphic computing. Our work will promote the study of FiM-based devices.

†Corresponding Author: zhuzhf@shanghaitech.edu.cn

*These authors contributed equally to this work.

The data that support the findings of this study are available from the corresponding author upon reasonable request.

**Acknowledgements:** This work was supported by National Key R&D Program of China (Grant No. 2022YFB4401700), Shanghai Sailing Program (Grant No. 20YF1430400) and National Natural Science Foundation of China (Grants No.

## APPENDIX A: The relationship between overdrive and the logarithm of the pulse width

In FM systems, in the precessional region, and the relation between overdrive $i$ and pulse width $\tau$ is given by[19] $i = 1 + \frac{\ln(\pi/2\theta_0)}{\tau/t_0}$, i.e., $i$-$1/\tau$ has a linear relationship. In other words, when the FM system is not affected by thermal fluctuations, $i$-$1/\tau$ becomes linear. However, in the thermally activated region, FM fellows $i = \left(1 - \frac{1}{\Delta}\ln\frac{\tau}{\tau_0}\right)$, i.e., overdrive $i$ is proportional to the logarithm of the pulse width. As shown in Fig. 3(b), the $i$-$1/\tau$ curve of S1 deviates from linearity under small $\tau$, we attribute it to the effect of thermal fluctuations. To verify our hypothesis, as shown in Fig. 5 we plot the curves of overload $i$ versus logarithm of pulse width for S1 and S2. The results show that the $i$-$\lg(\tau)$ curve of S1 is close to a straight line even under small $\tau$, which proves that S1 is seriously affected by thermal fluctuations. On the contrary, the $i$ of S2 under the small pulse width is proportional to $1/\tau$, which proves that it is less affected by thermal fluctuations.

## APPENDIX B: The standard deviation of the net charge required to achieve the same $P_{SW}$

As shown in Fig. 4(a) and Fig. 4(b), the $P_{SW}$-net charge curves of S1 shift significantly under different $\tau$, while the curves of S2 overlap. Therefore, to compare the extent to which S1 and S2 are affected by thermal fluctuations, we calculated the standard deviation of the net charge required to achieve the same $P_{SW}$ for S1 and S2. The standard deviation of the net charge to achieve the same $P_{SW}$ for two systems between 10ps and 500ps is calculated in the following way, $\sigma_{P_{SW}} = \sqrt{\sum_{i=1}^{n}(x_{i,P_{SW}} - \overline{x_{P_{SW}}})^2/n}$, where pulse width $\tau$ has nine values: 10ps, 20ps, 30ps, 50ps, 100ps, 200ps, 300ps, 400ps, 500ps, and $n = 9$. Meanwhile, $x_{i,P_{SW}}$ represents the value of $(J_C - J_{C0})\tau$ under a certain $P_{SW}$ and pulse width, and $\overline{x_{P_{SW}}}$ denotes the average value of the net charge required to achieve the same $P_{SW}$ at different pulse widths. For example, $x_{1,0.5}$ represents the value of $(J_C - J_{C0})\tau$ corresponding to $P_{SW} = 0.5$ under $\tau$ =10ps, and $\overline{x_{0.6}}$ represents the average value of $(J_C - J_{C0})\tau$ corresponding to $P_{SW} = 0.6$ under different $\tau$. As shown in Fig. 6, the standard deviation of the net charge required to achieve the same $P_{SW}$ for S1 is much larger than that of S2. These results further demonstrate that S1 is more sensitive to thermal fluctuations than S2.

**APPENDIX C: The net charge required for various $P_{SW}$ for FM systems**

For comparison, the dynamics of the FM/MgO/FM MTJ systems under thermal fluctuations are studied. The device structure is shown in the inset of Fig. 7(a). The parameters used in the FM systems are summarized in Table II, whcih are normal parameters in FM systems.[55] The anisotropy in this system consists of interfacial and bulk anisotropy with the anisotropy constant $K_u = K_b + K_i/t_{FL}$, where the bulk anisotropy $K_b = 2.245 \times 10^5$ J/m$^3$, the interfacial anisotropy $K_i = 1.286 \times 10^{-3}$ J/m$^2$. The effective anisotropy field is calculated by $\mathbf{H}_{k,\text{eff}} = (2K_u/M_s - 4\pi N_z M_s)m_z\hat{\mathbf{z}}$ with the demagnetizing tensors ($N_z$) calculated based on the sample dimensions. In order to ensure that the MTJ has perpendicular magnetic anisotropy, the free thickness $t_{FL}$ is set to be less than 1.5nm, because if it exceeds 1.5nm, the anisotropy will become in-plane. Therefore, to obtain FM system with different anisotropy, we set $t_{FL}$=0.9nm and $t_{FL}$=1.4nm for FM system 1 (FMS1) and FM system 2 (FMS2), respectively. Meanwhile, the ratio of the anisotropy of FMS1 to that of FMS3 is equal to the anisotropy ratio of FiM S1 and S2.

As shown in the Fig. 7(a), Fig. 7(b) and Fig. 7(c), the results demonstrate that the curves of net charge and $P_{SW}$ of the three FM systems overlap with each other under different pulse widths, respectively, indicating $P_{SW}$ is only determined by the net charge. Additionally, the standard deviation of the net charge required to achieve the same $P_{SW}$ for FMS1, FMS2 and FMS3 are measured, as shown in Fig. 7(d), they have similar and small standard deviations, which further demonstrates that the three systems are immune to thermal fluctuations at short $\tau$. The results indicate that changing the anisotropy does not lead to a change in the sensitivity of FM to thermal fluctuations. The change in sensitivity to thermal fluctuations is unique to the FiM systems. We attribute this anomaly to the complex relationship between anisotropy and overdrive in the FiM systems.

**Table II.** The parameters used in the FM systems

| Parameters | FMS1 | FMS2 | FMS3 |
|---|---|---|---|
| $\alpha$ | 0.0122 | 0.0122 | 0.0122 |
| $M_S$ (kA/m) | 1257 | 1257 | 1257 |
| Dimensions of FL (nm) | 50×50×0.9 | 50×50×1.4 | 50×50×0.9 |
| $\mathbf{H}_{k,eff}$ (T) | [0, 0, 1.13599] | [0, 0, 0.359711] | [0, 0, 0.037866] |

**APPENDIX D: The net STT required for 50% $P_{SW}$ of S1 and S2**

The extent to which the FiM system is affected by thermal fluctuations depends on the relative magnitudes of thermal fluctuations and STT. Only the STT overcomes the hindrance from anisotropy, and the remaining part is used to drive the magnetic moment motion, which is called the net STT. Meanwhile, S1 and S2 have the same normal distribution of the thermal fluctuation random field. In this case, we can compare the net STT of the two systems to determine their sensitivity to thermal fluctuations. Therefore, we compared the net STT at 50% $P_{SW}$ in S1 and S2, the net STT amplitudes of Fe at 50% $P_{SW}$ is represented by $\tau_{\text{net,Fe}} = \gamma_1 [\eta \hbar (J_{C,50} - J_{C0})]/(2t_{FL} M_{s,\text{Fe}})$.

As shown in Fig. 8, when the pulse width is very small, both systems have large net STT. At this time, the influence of thermal fluctuations on the two systems can be ignored. As the pulse width increases, the net STT of both S1 and S2 decreases dramatically. The effect of thermal fluctuations will gradually become comparable to the effect of STT, and the influence of thermal fluctuations can no longer be ignored. It is worth noting that with the increase of pulse width, the net STT of S2 will be significantly greater than that of S1. Since the effect of thermal fluctuations on the system depends on the relative magnitude of STT and thermal fluctuations, for two systems with the same normal distribution of thermal fluctuations, a smaller net STT means that S1 is more susceptible to thermal fluctuations than S2 with a larger net STT.

In addition, it should be noted that the analysis of "net STT" provides a feasible way to quantify the extent to which different systems are affected by thermal fluctuations. If this analysis is introduced to some other systems,[36] it can also help to understand the switching characteristics under thermal fluctuations from another perspective. However, it is not applicable to the analysis related to the switching process. The complex physical mechanism and properties of FiM require further investigation.

**Figures**

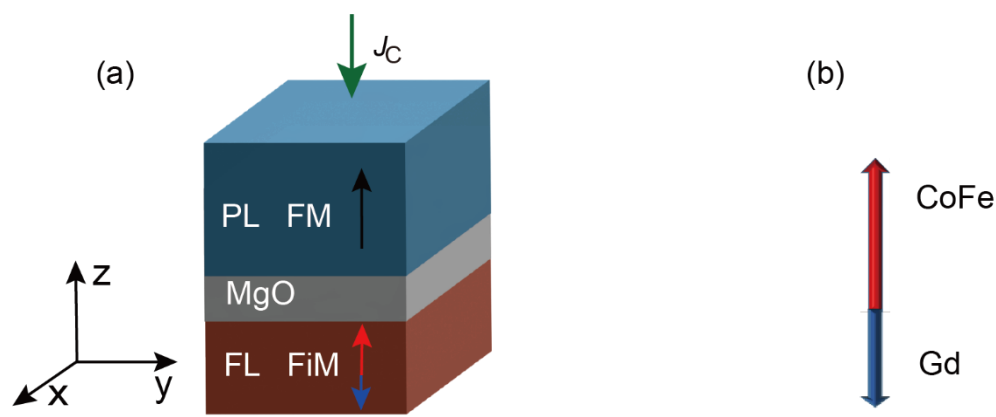

**Figure 1**. Illustrations of (a) the device structure and (b) the two-sublattice macrospin model.

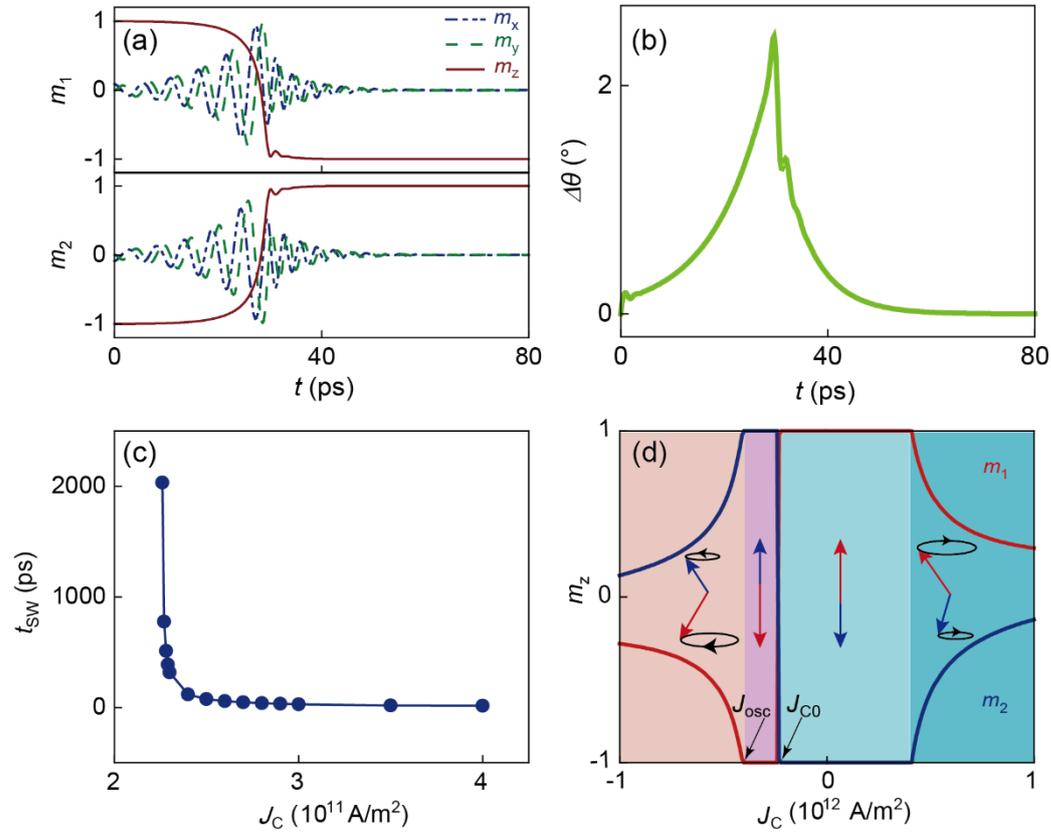

**Figure 2**. (a) Time evolution of $\mathbf{m}_1$ and $\mathbf{m}_2$. (b) The tilt angle by which $\mathbf{m}_1$ and $\mathbf{m}_2$ deviate from antiparallel (180°). (c) The relationship between switching time and current density. (d) The phase diagram of FiM with current density less than $1\times10^{12}$ A/m$^2$. The initial angle between $\mathbf{m}$ and the easy axis is set to be 5°.

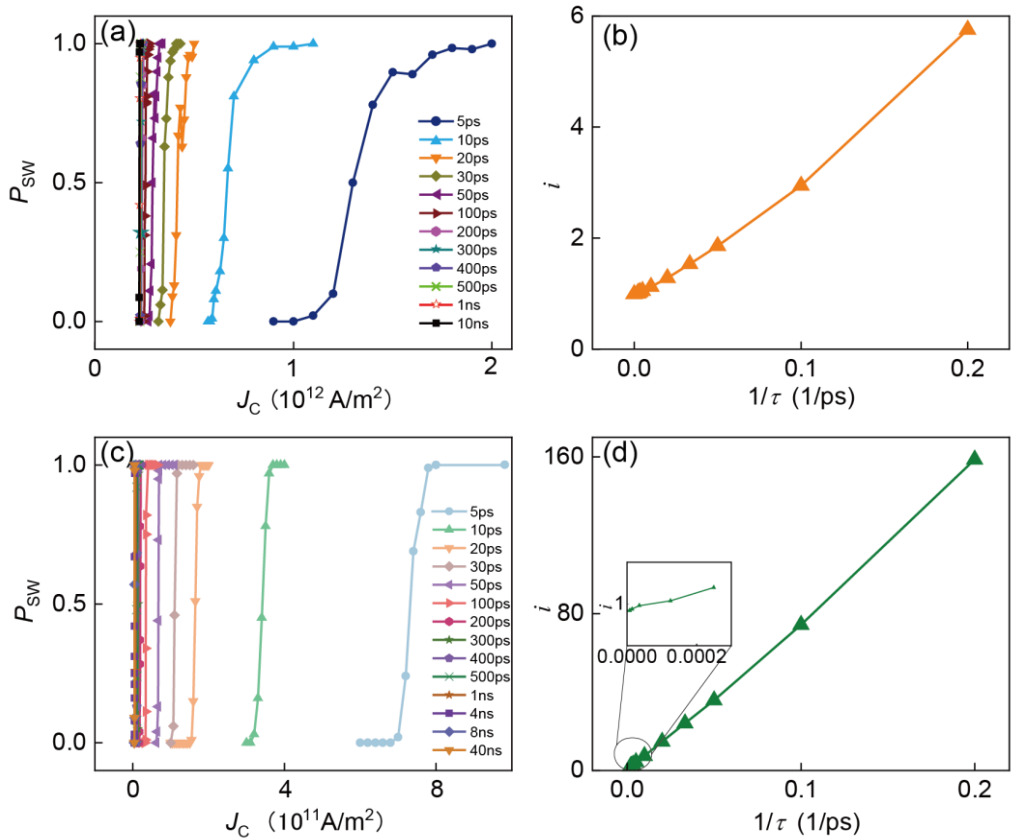

**Figure 3**. Relationship between switching probability and current density under different pulse widths of (a) S1. The overdrive $i$ as a function of the inverse of pulse width in (b) S1. Switching probability versus current density in (c) S2. The overdrive $i$ as a function of the inverse of the pulse width in (d) S2. When the effect of thermal fluctuations is considered here, the initial angle between **m** and the easy axis is set to 0° to ensure a thermally initial distribution.

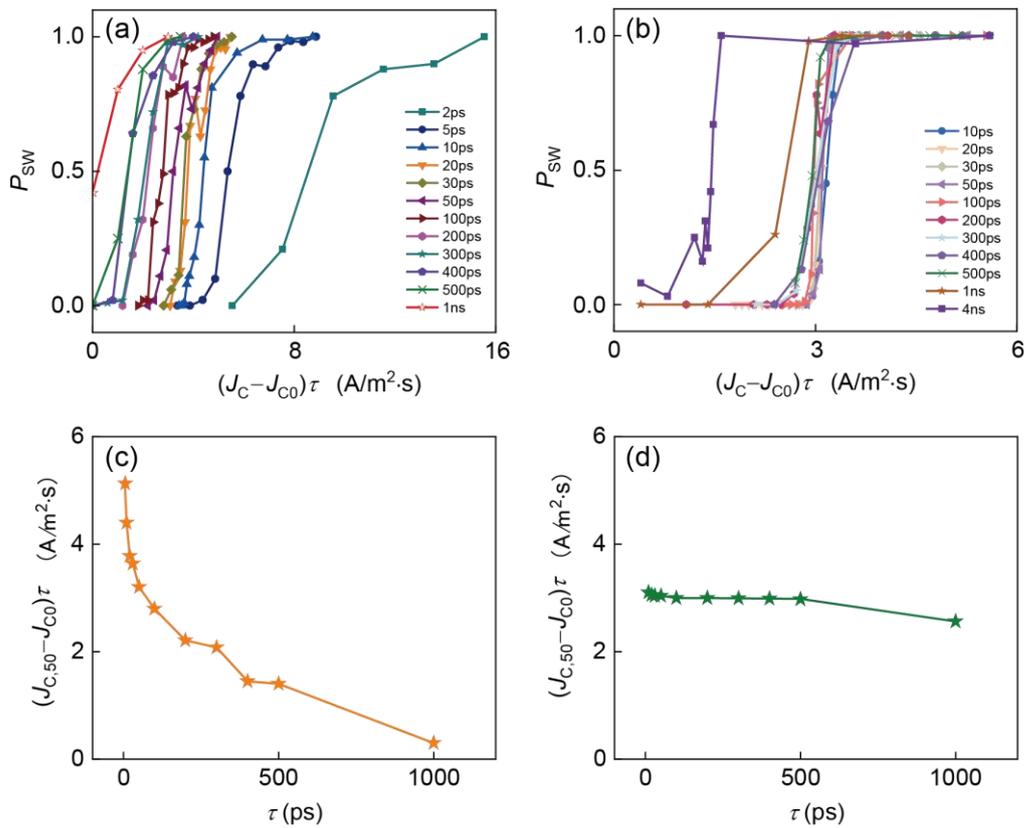

**Figure 4**. Switching probability versus net charge for different pulse widths in (a) S1 and (b) S2. The net charge corresponding to 50% switching probability as a function of pulse width for (c) S1 and (d) S2.

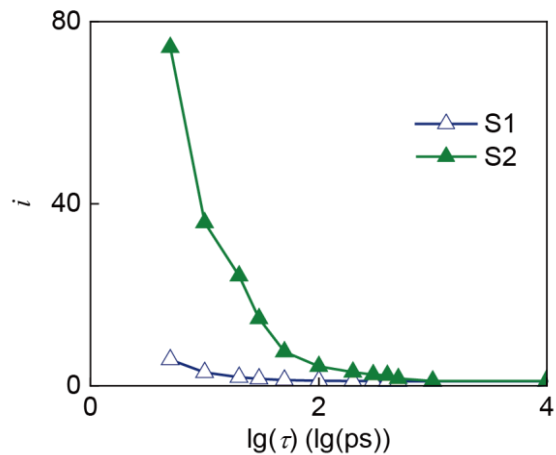

**Figure 5.** The overdrive $i$ versus the logarithm of pulse width in S1 and S2.

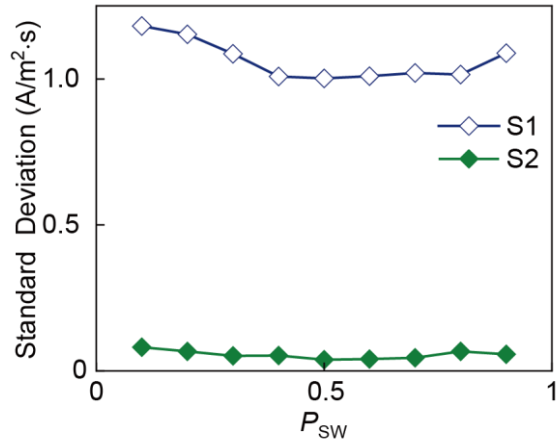

Figure 6. The standard deviation of the net charge required to achieve the same $P_{SW}$ for S1 and S2.

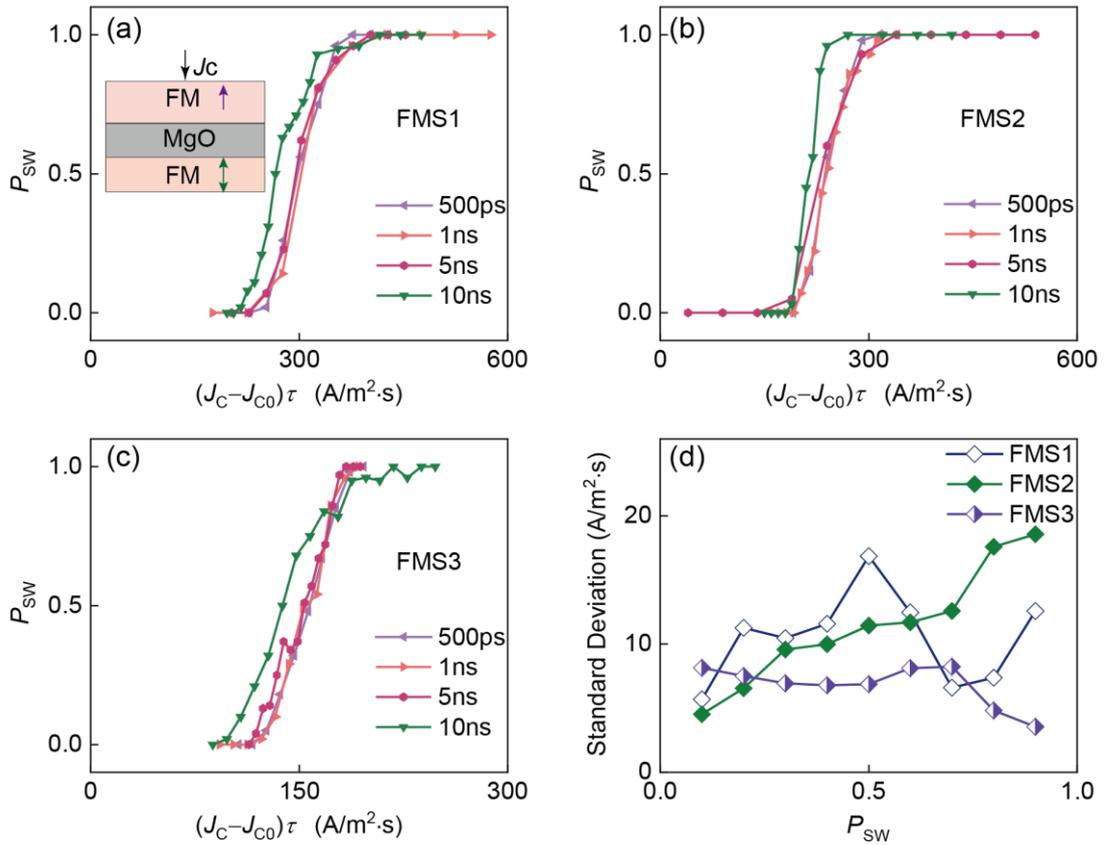

Figure 7. The net charge required for various $P_{SW}$ for (a) FMS1, (b) FMS2 and (c) FMS3 under different pulse widths. (d) The standard deviation of the net charge required to achieve the same $P_{SW}$ for FMS1, FMS2 and FMS3.

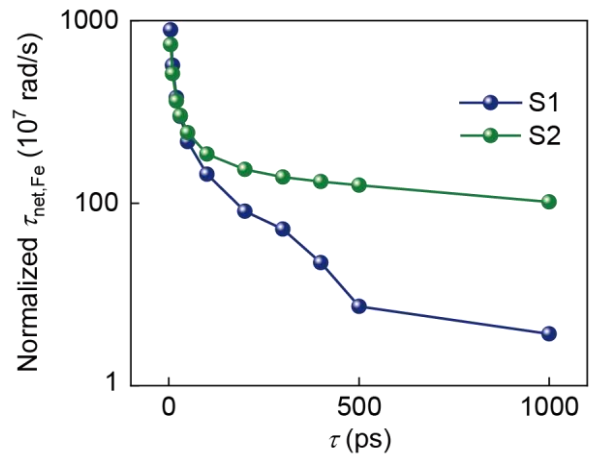

**Figure 8.** The net spin-transfer torque of Fe at 50% switching probability of S1 and S2.